# Prediction of Cytochrome P450-Mediated Metabolism Using a Combination of QSAR Derived Reactivity and Induced Fit Docking

Shulu Feng*,[†], Richard A. Friesner[‡]

[†] Schrödinger Inc. 120 West 45th street, New York, New York, 10036, United States
[‡] Department of Chemistry, Columbia University, New York, New York, 10036, United States
E-mail: shulu.feng@schrodinger.com

## Abstract

Prediction of metabolism in cytochrome P450s remains to be a crucial yet challenging topic in discovering and designing drugs, agrochemicals and nutritional supplements. The problem is challenging because the rate of P450 metabolism depends upon both the intrinsic chemical reactivity of the site and the protein-ligand geometry that is energetically accessible in the active site of a given P450 isozyme. We have addressed this problem using a two-level screening system. The first level implements an empirical QSAR-based scoring function employing the local chemical motifs to characterize the intrinsic reactivity. The second level uses molecular docking and molecular mechanics to account for the geometrical effects, including induced-fit effects in the protein which can be very important in P450 interactions with ligands. This approach has achieved high accuracy for both the P450 3A4 and 2D6 isoforms. In identifying at least one metabolic site in the top two ranked positions, the prediction rate can reach as high as 92.7% for the test set of isoform 3A4. For the 2D6 isoform, 100% accuracy is achieved on this basic evaluation metric, and, because this active site is considerably smaller and more selective than 3A4, very high precision is attained for full prediction of all metabolic sites. The method also requires considerably less CPU time than our previous efforts, which involved a large number of expensive simulations for each ligand to be evaluated. After screening using the empirical score function, only a few best candidates are left for each ligand, making the number of necessary estimations in the second level very small, which significantly reduces the computation time.

## 1. Introduction

Metabolism in P450 has crucial impacts on the bioactivity and the safety profiles of drug candidates. On one hand, P450 can convert compounds into a modified form that may interact differently with the drug target. On the other hand, it can also convert compounds into inactive excretable metabolites. Furthermore, P450 metabolism may lead to toxicity from unwanted products, causing failures in drug development. It is also relevant to many other important issues such as food-drug, drug-drug interaction and personalized medication.[1-3]

Predicting the drug metabolism in P450 is a very challenging topic for the following reasons. The first is due to the structural diversity of the P450 isoforms. The active site size varies considerably in different isoforms, e.g 3A4 (1386$Å^3$)[4] and 2D6 (~540 $Å^3$)[5]. The corresponding binding motifs are also quite different, e.g. in 2D6, the GLU216 and SER304 in the great majority of relevant cases forms a salt bridge with a positively charged nitrogen of the ligand, whereas 3A4 binds a wide range of ligands, and has no one feature of the active site that provides substantial geometric restriction. The second is that there are many possible reaction pathways that can be followed during the P450-catalyzed oxidation of the reactant. A recent review summarized the possible pathway of clinically used drugs which can be metabolized in the P450 1, 2 and 3 families.[6] A total of 248 pathways were proposed. One pathway can happen in several isoforms, which leads to some isoforms having as high as 148 possible pathways for the clinically used drugs, such as 3A4. Thirdly, there are certain reaction pathways that involving electron transfer in conjugated groups, which makes prediction even more difficult.[7]

Due to the challenges and resource requirement to study the metabolism experimentally[6], substantial efforts have been devoted to develop *in silico* method to study P450 metabolism[8, 9], such as the commercially available programs Metasite[10] and StarDrop[11]. Depending upon the input information that is to be utilized, there are generally three types of computational approaches: ligand-based, structure-based and mixed approaches.

Ligand-based methods makes predictions based on estimated hydrogen abstraction energies [12, 13]and they are usually combined with a heuristic attempt to take into account the ability of the ligand to bind to the P450 active site. Enzyme information may be incorporated to get improved results.[14, 15]. If a large training set is used, these methods can yield respectable results.[13, 15, 16] For example, RegioSelectivity-Predictor(RS-Predictor)[17, 18] uses topological and quantum chemical descriptors to represent the reactivity of potential metabolic site. However, due to the highly approximate consideration of the protein-ligand binding process, the precision and robustness of these approaches are limited, and it is difficult to increase accuracy beyond a given point.

On the other hand, structure based methods generally perform a thorough conformational search to generate the best possible poses, using molecular docking programs such as Glide[19], AutoDock[20] etc. Induced fit effects, that is, taking into account protein flexibility, are usually needed in such an approach to generate good results for mammalian P450 isozymes, which in general have a considerable degree of receptor flexibility upon ligand binding. However, the use of single protein structure in conjunction with a rigid receptor docking algorithm may leads to an inability to fully incorporate induced fit effects. Ensemble docking[21] can mitigate the problem of single protein structure by allowing docking a single ligand library against multiple rigid receptor conformations. Considerable evidence show that ensemble docking can improve the docking results. The improvement depends on the methods used to create the ensemble, the ensemble size and the scoring functions used in docking.

Mixed methods use both the reactivity calculation and conformational search to predict the site of metabolism. One way of combining reactivity information and structural information would be designing a scoring function which contains both the binding energy and the intrinsic reactivity. Highly accurate results can be generated in certain isoforms, such as our previous attempt on 2D6.[22] However, using a single scoring function requires highly accurate calculation of both the binding energy and intrinsic reactivity. Thus, the computational time may pose a problem. At the same time, an accurate estimation of the intrinsic reactivity would be a challenging task itself for the ligands where multiple reaction pathways would take place on certain sites. Finally, for 3A4, the size and complexity of the active site poses a severe challenge to a heavily physics based method. The incorporation of some empirical information based on a training set, while possibly reducing the generality of the methodology, enables the use of less demanding simulations, which is essential in bringing the CPU time for 3A4 calculations into the realm of practical applications.

The method described in the present work combines both ligand and structure based methods to generate an effective prediction approach. To mitigate the problem of computation time and the problem of multiple reaction pathways, a two-stage screening system is employed. The key is to locate a site of the ligand which possesses good reactivity and has an acceptable structure based binding energy in a suitably chosen induced fit pose. Considering the computational time required for accurate calculation of binding energies, the present work uses the reactivity as the first stage of screening. Unlike the traditional physics based methods of estimating the intrinsic reactivity, which attempt to calculate the reaction barrier of the ligand oxidation using for example quantum chemistry, the current method divides the candidates into different reaction groups with each group having similar reactivity. Only the candidates in the sufficiently highly reactive groups are passed on to the stage where relative binding energy is estimated, thus substantially reducing the required computational effort. As is shown below, this method achieves both high accuracy and high efficiency.

The paper is organized as following: the detailed methodology is provided in section 2; section 3 focuses on the results for 3A4 and 2D6; in section 4, the key functional groups for intrinsic reactivity, and induced fit effects are discussed; section 5 gives the conclusions and the directions of future work. We compare our 2D6 results to our own previous work [22] and demonstrate that the new approach achieves very similar levels of accuracy (including a false positive rate that is at least an order of magnitude better than alternative approaches in the literature) while reducing computation time substantially. We further investigate 3A4 because of its central importance in drug metabolism (~60% of metabolism is due to 3A4 for known drugs), and because of the great challenge of treating its large, exceptionally flexible active site; there was a real possibility that structure based methods simply would not be helpful for 3A4 metabolic predictions. We show that this is not the case, and that the induced fit docking step improves accuracy by ~10-15%; not as dramatically as for 2D6, but enough to make a measurable, and useful, difference. Furthermore, the present approach provides structures (which using present technology cannot be obtained experimentally) which may have utility in lead optimization efforts when there is a metabolic problem with a drug candidate.

## 2. Methods

The present work uses a series of hierarchical filters to search for the best sites of metabolism of the ligand. In the previous combined method to predict the site of metabolism in P450 that we developed, prediction of intrinsic reactivity and binding energy are both required to maintain accuracy. The current approach is also based on the intrinsic reactivity and binding energy. However, instead of combining the binding energy with intrinsic reactivity (IR) directly to produce a single overall scoring function, as was done previously, our new method constructs a series of filtering processes based on the IR of each potential site and the corresponding binding energies.

In the first level of filtering, the IR of each potential site of the ligand is calculated using a very rapid empirical scoring approach, discussed below in some detail. In our previous work on 2D6, we employed an approximate quantum chemical methodology, which was reasonably successful. However, the extension of such an approach to 3A4 has potential pitfalls. As mentioned previously, there are many possible reactive pathways to be considered; the number of such pathways is multiplied in 3A4 where geometric restrictions are much less severe than they are in 2D6. Furthermore, some possible pathways, especially those involving conjugated chemical functional groups, pose additional challenging to the binding energy estimation. In those pathways, the closest site to the oxo may not be the final site that has been oxidized. Therefore, it is difficult to build a direct relationship between the docked pose and the potential site of metabolism.

Many computational efforts have been tried or proposed to overcome these issues.[8] In the present paper, our method uses a score function, which characterizes the IR based on the local chemical motif. This is based on the assumption that the local chemical motif is the main factor that decides which pathway the potential site will choose and also the main factor that characterizes the reaction barrier for the potential site. Therefore, a method to characterize the functional group in the local motif of each potential site of metabolism and an approach to characterize the potential contribution of each functional group are required. The local motif reactivity score ($local\_motif\_RS$) function is then designed to address the above two issues. In this model, the reactivity of the target site of metabolism (SOM) is solely decided by the local chemical motif of that target site. The local chemical motif is characterized by three classes of functional groups: the primary, the secondary and the correctional functional groups. The primary functional groups are used to describe the immediate environment of the target SOM atom, and it is directly associated with chemical features, such as number of connected hydrogen atoms, aromatic properties etc. The secondary functional groups are used to describe the chemical groups directly connected to the target SOM, such as positive charged nitrogen, carbonyl, and benzene et.al. The correctional functional groups are used to describe the additional functional groups connected to the target SOM or the secondary functional groups. The overall reactivity score $local\_motif\_RS$ of the target SOM is then calculated based on the following formula:

$$local\_motif\_RS = Primary\_RS + Secondary\_RS + Correction\_RS$$

$Primary\_RS$, $Secondary\_RS$ and $Correction\_RS$ are the reactivity score assigned to the primary, secondary and correctional functional groups of the target SOM.

All the functional groups included in the model are identified from the published experimental results and the proposed mechanisms[23-26]. For the primary functional groups, the element carbon (C), nitrogen (N), phosphorus (P) and sulfur (S) are considered as the potential SOM. Most of the experimental observed SOM has the element of C. Therefore, most of the primary functional groups are carbon based. The carbon based primary functional groups are then characterized by number of connected hydrogen atoms, whether the target atom is in aromatic ring or not and if the target atom is in certain ring structures. The nitrogen based primary functional groups are further divided based on number of connected hydrogen atoms and also if the target nitrogen is in aromatic structure or not. For the sulfur based primary functional groups, only sulfur with oxidation number less than +4 are considered as the potential SOM. These functional groups are only further divided depending on whether the sulfur is in an aromatic ring. For the phosphorus related functional groups, only phosphorus atoms connected with sulfur atom are considered as potential SOM.

All the secondary functional groups are constructed based on the potential reaction mechanisms for the target potential SOM. There are several reviews[26], which provide excellent summary of the possible reaction path for each type of chemical motif. In the current method, the secondary functional groups are divided into the following four categories: (1) columbic attraction type; (2) conjugated group; (3) lone pair group; (4) specific structure. The columbic attraction type mainly composes of positively charged nitrogen. Many experimental observed SOMs are near the positively charged nitrogen, for example, chloroquine[27], citalopram[28], clozapine[29] et.al. One explanation of this could be that the positive charge on the nitrogen can form a salt bridge like structure with the negatively charged oxo, therefore stabilizing the docked structure. However, the nitrogen atom may actually join the reaction in a neutral state and provide an electron to the oxo, such as N-dealkylation reactions. For this reason, all the nitrogen atoms, which are not in conjugated systems, are considered to belong to the same functional group. The conjugated groups are the most complicated type of chemical groups. This category can facilitate the reaction by help the electron transfer process through conjugated pi bond. For example, the primary metabolic site of testosterone [30] is connected with a carbon double bond, which is further conjugated with a carbonyl group. The lone pair groups of the carbonyl can help the carbon form a stable radical, therefore facilitating the radical based reactions. For example, the true positive site of BDB[31] has two oxygen atoms connected. The last type of group is the chemical motifs with specific structures. Most of these specific structures consist of a ring structure, e.g. 1,8-cineole[32]. The detailed information on how this kind of structure may affect the oxidation process is unclear. In the current study, there is only a few of this type of functional groups considered.

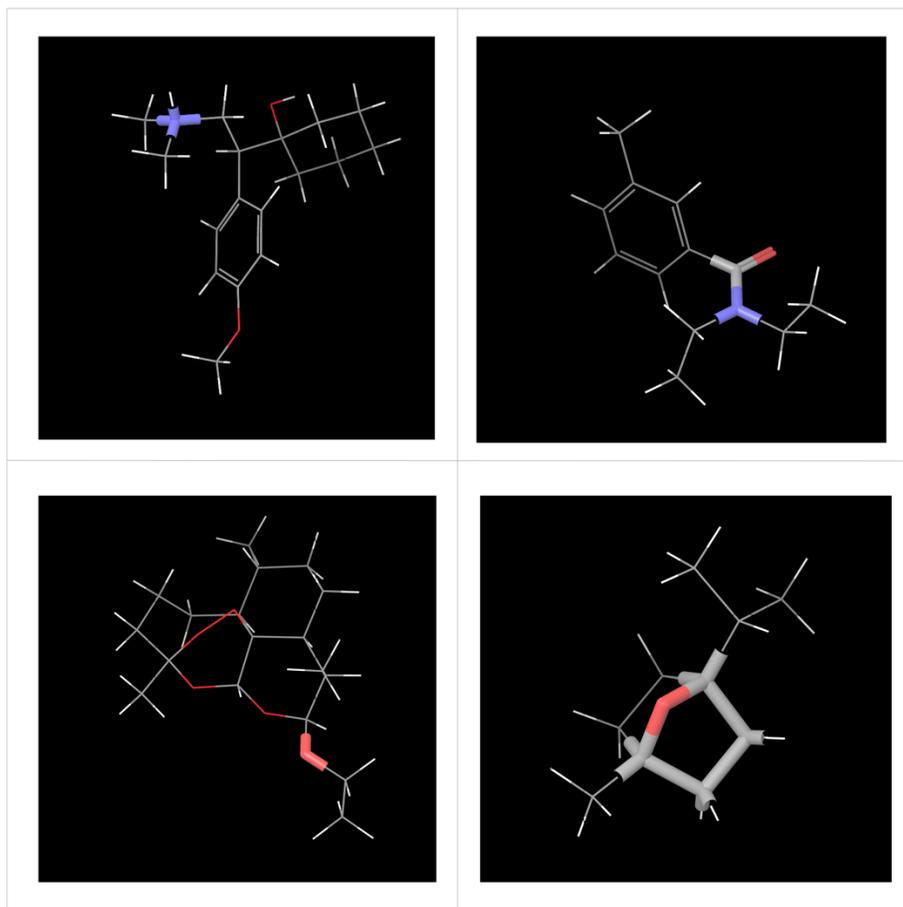

Figure 1: Examples of the different type of secondary groups. The color scheme is: red(oxygen), white(hydrogen), blue(nitrogen), grey(carbon). All the groups are highlighted in tube. Upper left: the columbic attraction type, a positive charged nitrogen in venlafaxine is highlighted; Upper right: the conjugated type, organic amide group in NN_dimethyl_M_toluamide is shown; lower left: the lone pair type, an oxygen in the beta_arteether is shown; lower right: the specific structure type, a five-member ring is highlighted.

The correctional groups are similar to the secondary groups. They are added to describe the target atom mainly in the following two conditions, (1) the target atom connected to more than one functional group; (2) the functional group is conjugated with secondary functional groups. The only exception to the above two conditions is that, when a nitrogen related functional group is taken as the secondary group, then the nearby oxygen atom within 2 bonds will be considered as a correctional group. For the first condition, if more than two functional groups are directly connected, only the two with the top two strongest effects are considered.

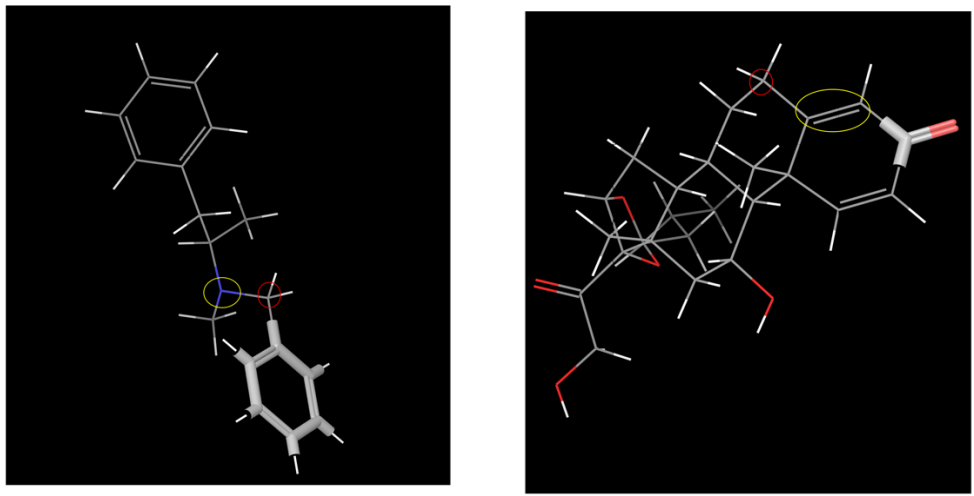

Figure 2: Example of the correctional groups. Red circle: the target atom. Yellow circle: the secondary group. Tube: the correctional group. Left: the correctional group is directly connected with the target atom in benzphetamine. Right: the correctional group is conjugated with the secondary group in budesonide.

Only the moieties that are believed to be able to substantially change the reactivity of the target SOM are considered. When constructing the groups, we tried to use small functional groups or break the large functional groups into small pieces with the aim of increasing transferability and generality. This would leads to some kinds of unwanted combinations. Therefore, not all combinations of different types of primary, secondary and correctional functional groups are considered. All of the groups considered in the case of 3A4 are summarized in the supplemental material.

The groups are further organized into 16 classes. For all the groups in the same class, identical score is assigned. Therefore, in this method, there are only 16 parameters needed. In addition, since only the ranking is important, each parameter is designed to pick a level from 40 levels. Thus, this method has a very small solution space for the parameters, helping to reduce potential over-fitting. One more effort to reduce overfitting is that we assign, prior to numerical optimization, a high score to certain groups, which are highly reactive, and do not change this value during the optimization process.

In the second stage, all the candidates in the top 2 categories from the first stage are further refined. This refinement process requires molecular docking, geometric sampling of the protein active site side chains, and minimization of the protein-ligand complex, to account for the induced fit effects. In the molecular docking stage, constraints were added to make sure that there are reasonable candidates in the neighborhood of the iron-oxo complex. Two types of constraints were used to match the needs of different type of isoforms. For the isoforms with large active site size, such as 3A4, a constraint that specifies each possible candidate SOM (by constraining the distance of the target atom to the ferryl oxygen) is used. The purpose here is to make sure that each possible candidate will have a corresponding geometric structure that can goes to the geometric optimization. The reason is that a large number of possible pose can be generated, given the large size of the active site. Due to this reason, it is difficult to set an arbitrary cut-off, e.g. number of poses, which will ensure all the candidates can have a corresponding structure. For the isoform with relatively small active site sizes, such as 2D6, a general constraint that only requires one heavy atom in the neighborhood of the iron-oxo is used. The consideration behinds this is that it will require a great deal of computational time to fulfill the first type of constraints in a small active site.

For the structural optimization, the Protein Local Optimization Program (PLOP, known as Prime in Schrödinger Package)[33] software package is used. The optimization begins with a structural minimization followed by hybrid Monte Carlo sampling. The structural minimization will minimize the ligand as well as the active site region of the protein. The hybrid MC sampling consists of multiple parallel simulations, as in our prior methodology.[22] In that work, three types of randomized movements were considered: side-chain torsional rotations, rigid-body movements of the ligand, and molecular dynamics trajectories, all using the continuum solvation model in PLOP. In the present work, the rigid body movements are not used due to its low acceptance ratio and relatively long sampling time. After the first stage screening, only a small number of candidates are selected to go through the second stage of screening. Computational experiments have shown that, using our new filtering algorithm, less extensive simulations can be employed in the structural optimization component of the algorithm (as well as fewer simulations being required). These modifications to the algorithm results in a dramatic decrease in computation time per ligand, as compared to our prior approach.

The side-chain movements are used to sample the selected side-chain conformation of the protein and the ligands by varying the dihedral angles of the rotatable bonds. In each movement; up to three close residues are allowed to rotate their side chains. The new rotational conformations are constructed through random perturbation or pre-defined rotamer libraries. If the random perturbation is used, the change of each dihedral angle is the sum of a large rotation and a small rotation. The large rotation is N*60 degrees in which N is a random integer from 0 to 5. The small rotation is a random number from 0 to 30 degrees. If the rotamer library is used, a random rotamer from a high resolution side-chain library for protein residues [34] and from a library of 10 degree resolution for ligand are used to create a new conformation. The new structure, with acceptably low van der Waals clashes with other protein atoms, is then minimized and subjected to the acceptance criteria.

The hybrid MC[35] movements sample the conformation space of the selected residues in the protein and the ligand. A 5 picosecond, constant energy molecular dynamic (MD) simulation is performed to generate the new structure. In the MD simulation, the RESPA based integration of short range forces is used with a time step of 1 femtosecond while the long range forces are updated every five steps using Verlet integration[36].

In all the sampling steps, distance and angle constraints are applied to the target atom to make sure the target atom is in a reasonable region of the active site. Detailed information is shown in Figure 3 and 4.

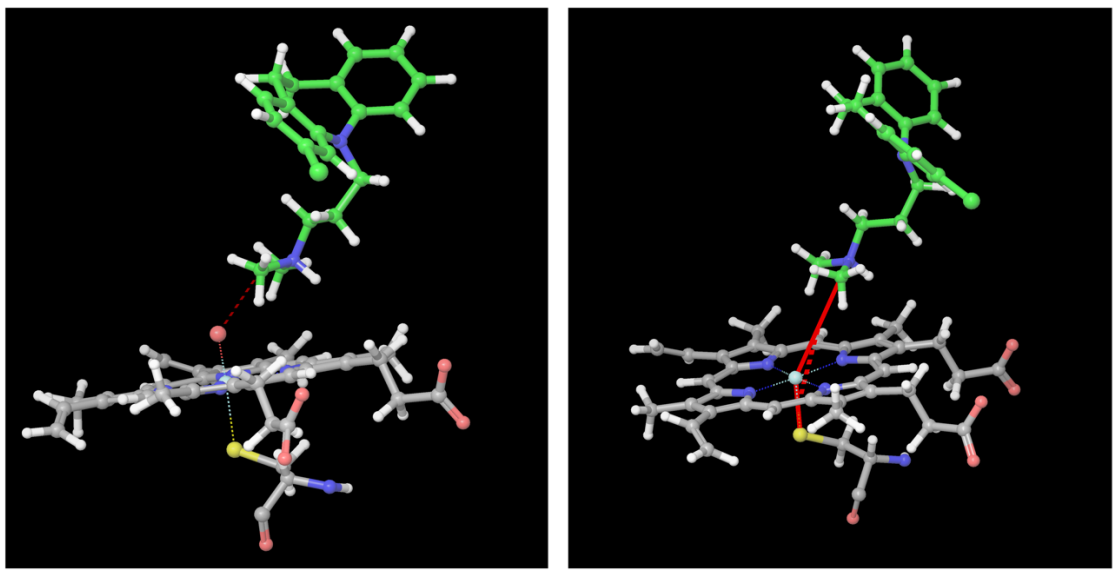

Figure 3: The constraint applied on the non-aromatic target. The color scheme is: pink(oxygen), white(hydrogen), yellow(sulfur), aquamarine(iron), green(carbon on the ligand), grey(carbon in the heme ring and protein). Left: the distance constraint (2.5±0.5Å) applied to the target atom and the virtual oxo, represented by the red dashed line. Right: the angle constraint (145±20°) applied to the target, represented by the red solid line.

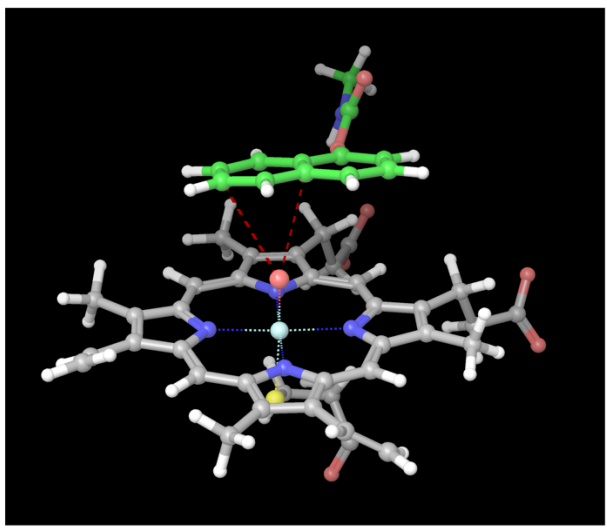

Figure 4: The additional distance constraints (2.8±0.5Å) applied on the aromatic target. The color scheme is the same as that in Figure 3. The two distance constraints are applied to the two nearby atoms, as represented by the red dashed line.

The final step in the current work is evaluation of the predictive power of the methodology, that is, agreement between experimental measurement of the SOM and the computational ranking of the various ligand sites .Two types of metrics are used in the evaluation process, the top metric (top 1 and 2) and the exact number of true positives (TP), false positives (FP) and false negatives (FN). We define the top 1 and top 2 metrics following Breneman and coworkers: the top 1 metric is the percentage of time that the best scoring site found by the model is experimentally validated as an SOM, while the top 2 metric is the percentage of time that one of the two best scoring sites has been experimentally validated as an SOM. Use of this metric enables direct comparison with the approach of Breneman and also other methods (MetaSite, Stardrop) that are reported in ref. [17, 18]. The second approach counts the number of predictions that agree and disagree with the experimental results, and thus is a comprehensive assessment of the accuracy and reliability of the methodology.

For the first type of metric, all the candidates are first sorted by the local reactivity score. Then, for the candidates with the same local reactivity score, they are further sorted based on the energies calculated from the structural optimization stage.

For the second type of metric, the evaluation stage is constructed by using cut-off on both the reactivity score and energy of the structure. All the top 2 categories of atoms are considered as potential site of metabolism. Then a rejection based method is used to get rid of the bad candidates. For the top-1 category of atoms from the reactivity score, if the corresponding energy is more than 50 kcal/mol higher than the lowest energy, this atom is rejected. For the atoms in the second category, if the corresponding energy is more than 20 kcal/mol higher than the lowest energy or the reactivity score is smaller than 1, this atom is rejected as well.

The second type of metric is used when the first type of metric is not capable of distinguishing the difference of the results from different methods or stages, i.e. in the case where the top2 metric reach almost 100%. In the current simulation, due to the diversity and complexity of 3A4, only the first metric is used in the prediction of 3A4. For 2D6, since the top2 result of the training set already reaches an accuracy of 100%, both the first and second metric are evaluated.

Put another way, we cannot expect the current sampling algorithm to be able to rigorously sampling the phase space of 3A4; the active site is too large, there are too many side chains that can adopt multiple rotamer conformations, and there is no anchor point, as there is in 2D6 to constrain the ligand. Our computational approach, and evaluation metrics, are designed to recognize this reality. In contrast, 2D6 has a much smaller active site, and an anchor point in the salt bridge formation of a positive nitrogen on the ligand to GLU216/SER304. These factors drastically reduce the phase space available to any given ligand, and enable the induced fit methodology to provide a very large improvement in the accuracy of prediction for all sites on the molecule, including secondary (and less active) metabolic sites. Further work on both methodology and analysis of ligand binding modes for 3A4 will facilitate progress in moving towards results of the quality of those obtained here for 2D6.

### *Application to different isoforms*

Despite the structural diversity of different isoforms, it is not optimal to construct from scratch a unique library of functional moieties for each isoform. Rather, the application to isoforms other than 3A4 will take the functional moieties library constructed for 3A4 as the base. Necessary modifications which account for the isoform related binding motifs can be added to the method used in 3A4. There are two considerations in deploying this kind of base-extension system. The first is because the underlying reaction is similar. Secondly, the data sets are small for certain isoforms. We choose 3A4 as the base because of (1) the large number of possible reaction pathways; (2) its large active site, which makes the training process for building the intrinsic reactivity model less affected by the protein side chains, thus the functional moieties can better represent the intrinsic reactivity and have better transferability. In the present study, 2D6 has chosen as the isoform to test the extension part of the current methodology. There are two types of modifications in addition to the original base method. The first type is the modification that accounts for the salt bridge formation between GLU216/SER304 and the positive charged nitrogen on the ligand. The second type of modification is used to adjust the functional moieties. This type of modification is learned from the reaction pathway difference between 2D6 and the base isoform 3A4.

## 3. Results

*3A4*

The data set used for the 3A4 case contains 384 ligands. This data set is basically the same as the one used in Breneman's work[17], except that large ligands, which contain more than 60 heavy atoms, are not considered in current study, due to the difficulty to generate reasonable poses in induced fit docking. Modification of the induced fit docking algorithm to handle larger systems is possible and is planned in future work. For now, the vast majority of pharmaceutically relevant compounds have fewer than 60 heavy atoms, and it is reasonable to ask what the results are for the data set when such cases are eliminated.

The PDB structure used for 3A4 is 3UA1 [37]. The overall results are summarized in Tables 1 and 2. In Table 2, 178 ligands are randomly chosen as the training set. The top 2 result is taken as the target during the parameter fitting process. Experiments with different random seeds, which generate different training sets, produced minimal variation in the results, as indicated by the standard deviation shown in Table 2.

Table 1: The overall results of 3A4

| Method | TOP1 (%) | TOP2 (%) |
|---|---|---|
| RS Boltzmann [a] | 60.7 | 78.2 |
| SMARTCyp[a] | 63.4 | 73.2 |
| RS-Predictor(TOP QC SCR) [b] | 67.2 | 82.3 |
| Current Method | 73.5±1.6 | 93.2±1.1 |

(a) The data set[17] is the same as used in the current method except the extra-large ones. (b) The data set[18] is different than the current method by including ~90 more ligands, which contains the extra-large ones.

Table 2: The results of 3A4

|  | TOP2 (%) |
|---|---|
| Training | 93.4±1.2 |
| Test | 92.7±2.3 |
| Overall | 93.2±1.1 |

The results are compared with other method on the similar data set. Most of the extra ligands used in the RS-Predictor(TOP QC SCR)[18] method are not drug like molecules. Therefore, these additional molecules are not tested in current method. For the top 2 metric, the best result reported for alternative methodologies is from RS-predictor, which is based on the RS Boltzmann consensus model, and reaches 78.2%[17]. The result from our method shows a ~ 15% improvement over RS-Predictor. With respect to the top 1 metric, the best alternative performance is obtained by SMARTcyp[12], which reaches a rate of 64.6%. Our method can reach as high as 73.5%, which is ~ 9% better.

The average computational time for prediction SOM in 3A4 is less than 1 hour CPU time on a single 2.6 GHz AMD core. This is clearly unproblematic for late stage lead optimization applications (where a small number of compounds would be considered), and is viable for screening of tens or hundreds of thousands of compounds using a cluster of moderate computational capabilities.

## *2D6*

The data set used to test 2D6 is the same data set that was used in our previous effort, IDsite[22]. All the ligands in the data set are believed to interact with the protein via one or more salt bridges, as described previously in the method section. The results are summarized in Table 3. The training set and the test set are the same as the ones used in IDsite.

The parameters developed for the reactivity model from the 3A4 data set are used to characterize the functional groups. The training set is mainly used to adjust the group partitions and to fit the salt bridge motif. In both the top 1 and top 2 metrics, the current method gets very high prediction rates. The actual numbers of true positives (TP), false positives (FP) and false negatives (FN) are provided in Table 4 and 5. As noted above, this encompasses all sites of reactivity reported in the literature.

Table 3: The overall results of 2D6

| Metric | Current Method (%) | RS-WebPredictor (%) |
|---|---|---|
| TOP1 (training set) | 91.7 | 63.9 |
| TOP1 (test set) | 85 | 65 |
| TOP2 (training set) | 100 | 77.8 |
| TOP2 (test set) | 95 | 80 |

Table 4: The training set results of 2D6

| Method | TP | FN | FP |
|---|---|---|---|
| Current Method | 51 | 6 | 8 |
| IDsite | 47 | 10 | 13 |
| IDsite fitted | 52 | 5 | 8 |
| RS-TOP1 | 26 | 31 | 13 |
| RS-TOP2 | 35 | 22 | 49 |
| RS-TOP3 | 42 | 15 | 93 |

RS-TOP1, RS-TOP2 and RS-TOP3 represent the results from RS-WebPredictor by using the top1, top2 and top3 metric.

Table 5: The test set result of 2D6

| Method | TP | FN | FP |
|---|---|---|---|
| Current Method | 24 | 1 | 8 |
| IDsite | 21 | 4 | 8 |
| IDsite fitted | 25 | 0 | 6 |
| RS-TOP1 | 14 | 11 | 7 |
| RS-TOP2 | 21 | 4 | 26 |
| RS-TOP3 | 21 | 4 | 47 |

IDsite and IDsite fitted are from the previous work. RS-TOP1, RS-TOP2 and RS-TOP3 represent the same method, as described in Table 3.

To get a better understanding of our current results, the same data sets are used in RS-WebPredictor[38], which is the web service of RS-predictor. Table 4 and 5 summarized the results from IDsite, RS-WebPredictor and our method. In both tables, we can see that our current method gets similar results as the fitted IDsite does; these methods are close to precise reproduction of experiment. However, the present method has significantly reduced the computational time required. The average computational time for prediction SOM in 2D6 is less than 5 hour CPU time on a single 2.6 GHz AMD core. When comparing with the calculated results from the RS-WebPredictor, the present method is qualitatively better in terms of all the TP, FP and FN, by an order of magnitude or more depending upon exactly what metrics are used to make comparisons. These results demonstrate that, as docking accuracy increases, the power of an explicit structure based component of the model becomes much larger, and leads to highly reliable SOM identification.

***Results at different stages***

The induced fit effects for both 3A4 and 2D6 are summarized in Table 6 and 7. For 3A4, the prediction rates increase by more than 9 percent, giving that the prediction rate from the reactivity score function is already as high as 83.9%. For 2D6, the induced fit effects are also important, as shown in Table 7. The number of TP increase by more than 20% for both the training and test set.

Table 6: The results of 3A4 at different stage

| Stage | TOP1 | TOP2 |
|---|---|---|
| Without minimization | 65.9±1.1 | 83.9±0.4 |
| After minimization | 73.5±1.6 | 93.2±1.1 |

The results without minimization are calculated by using the local reactivity score only.

Table 7: The results of 2D6 at different stage

| Type | No induced Fit | Minimization | Sampling |
|---|---|---|---|
| TP(training) | 39 | 49 | 51 |
| FN(training) | 18 | 8 | 6 |
| FP(training) | 3 | 13 | 8 |
| TP(test) | 18 | 24 | 24 |
| FN(test) | 7 | 1 | 1 |
| FP(test) | 4 | 10 | 8 |

No induced Fit: results are calculated by using the local reactivity score only. Minimization: results are calculated by only using the energy from the minimization of the docked poses. Sampling: the results are calculated by using the full protocol as described in the method.

## 4. Discussions

*Functional groups*

Among all the functional groups used, there are several types that are the most important. Table 8 provides the top 2 most important functional groups in the secondary as well as the correctional types. One obvious feature is that they are all very simple groups. The nitrogen atom, which is not in the conjugated system, can help the oxidation process in different ways, as mentioned in the method section. The oxygen atoms mainly help the reaction by stabilizing the carbon radical through lone pair. The top 2 correctional groups are interesting. The first one is the methyl group. On one hand, the correctional methyl group is related to the dealkylation type of oxidations. On the other hand, a target connected with the methyl group is usually at a very flexible position of the ligand. Thus, it is more correlated with the positive reaction sites than negative reaction sites. The carbonyl group usually acts to reduce the positive effects of the secondary group through conjugation. Therefore, the correctional carbonyl group is more correlated with negative sites than positive ones.

Table 8: The top2 most important secondary and correctional functional groups

|  | Structure | Frequency(total 384) |
|---|---|---|
| First secondary | N (not in conjugated system) | 202 |
| Second secondary | -O- | 155 |
| First correctional | -CH3 | 137 |
| Second correctional | C=O | 106 |

## *Induced Fit effects*

Even though the intrinsic reactivity plays the most important role in the case of 3A4, the induced fit calculation is also indispensable to get a satisfactory prediction rate. For the current data site, about 116 ligands need to go through the induced fit calculation, which further improves the top 2 metric result from 83.9% to 93.2%. Figure 5(a) and (b) shows two minimized poses of arachidonic_acid docked into 3A4 structure (PDB code: 3UA1). By taking into consideration induced fit effects, the candidate structure in Figure 5 (a) has a more favorable conformation than the candidate in Figure 5(b). It is clear that the two strong salt bridges formed between the protein and the carbonate in Figure 5(a) play a significant role in selecting the correct pose. One is constructed with the backbone of PHE213 and the other is constructed with the side chain of ARG212. For the candidate in Figure 5(b), only one salt bridge is observed with GLU374. The distance between the oxygen and hydrogen is larger than the ones in Figure 5(a).

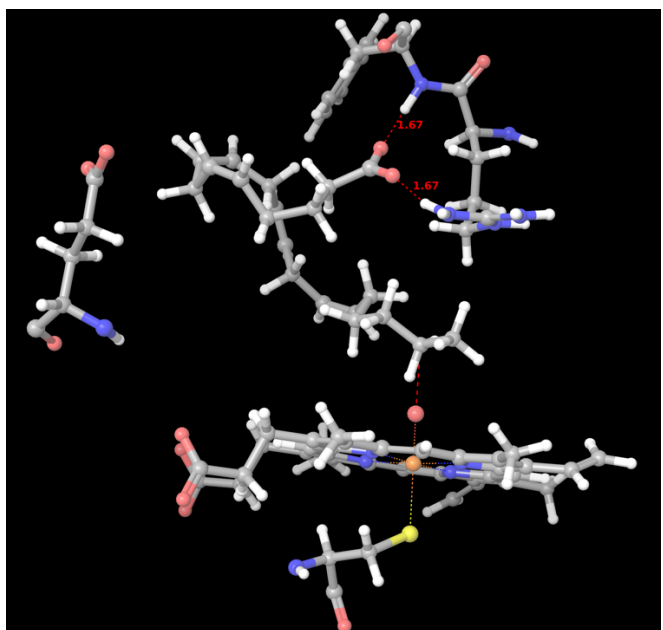

(a)

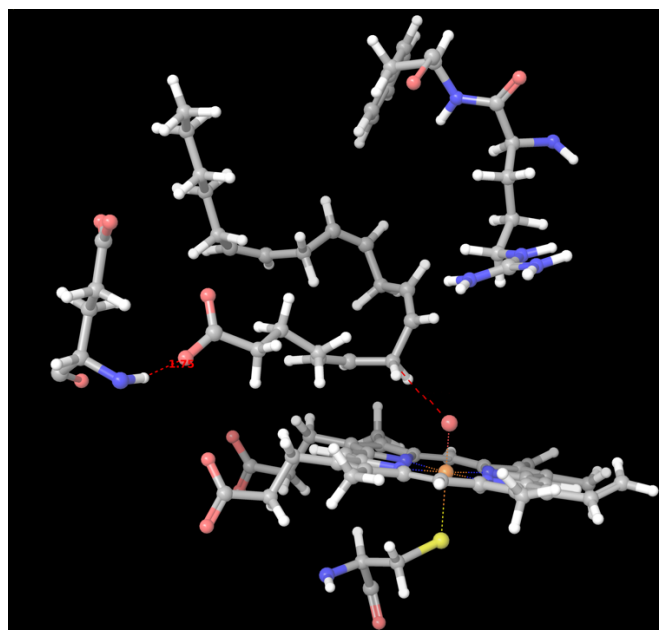

(b)

Figure 5: The docked pose of arachidonic_acid in P450 3A4. The color scheme is: pink(oxygen), white(hydrogen), yellow(sulfur), orange(iron), green(carbon). (a) pose for positive. (b) pose for negative. The red dotted line with number shows the connection of the formed salt bridge between the ligand and protein. The red dashed line without number connects the oxo and target atom of the ligand.

The importance of induced fit can be further analyzed by examining the actual conformational change of the sampled region. Figure 6 shows the structures before the minimization stage in 3A4 for the ligand arachidonic_acid. Comparing with minimized structure shows in Figure 5(a), it is clear that the protein underwent conformational change to better interact with the ligand, as indicated by the large oxygen-hydrogen distance changes in the salt bridge regions.

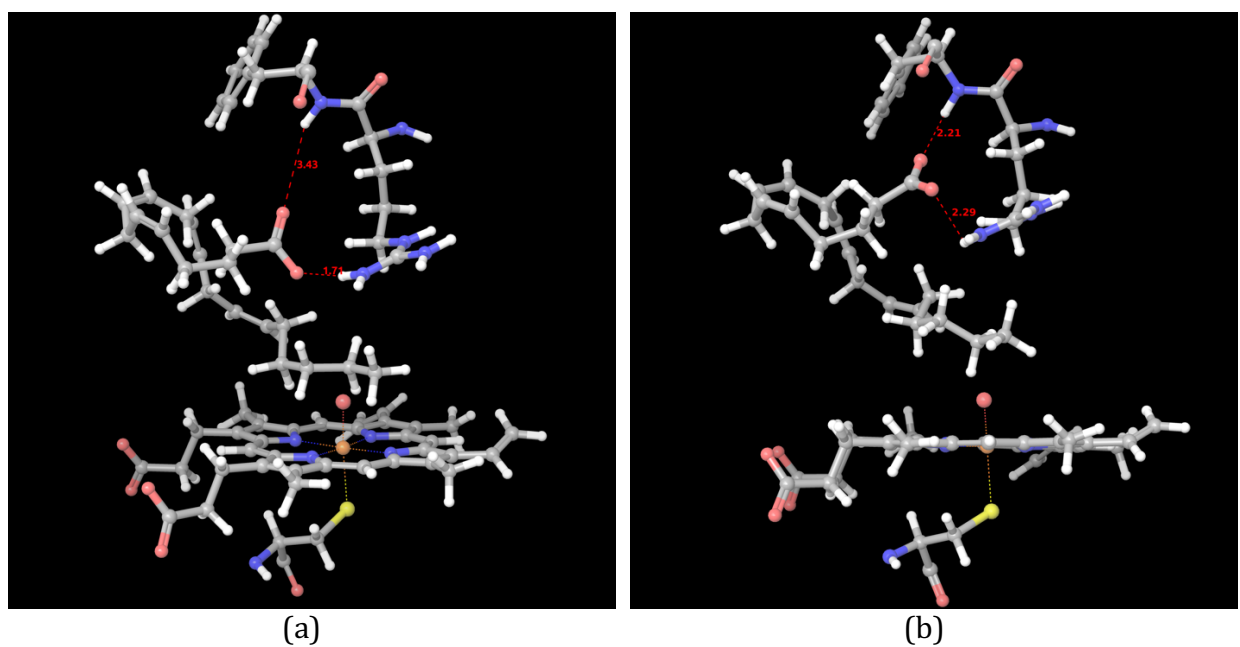

(a)                                    (b)

Figure 6: (a) The docked pose of Figure 5(a) before minimization. (b) Ligand structure after minimization combined with protein structure before minimization. The red dashed line with number shows the connection of the formed salt bridge between the ligand and protein. Color scheme is the same as that in Figure 5.

In the case of 2D6, the induced fit calculations have a larger impact on prediction accuracy due to the relatively small active site size. As show in Table 5, the number of TPs has increased from 68.4% to 89.5% for the training set, and from 72% to 96% for the test set. At the same time, the number of FPs has only increased by 5 for the training set and 4 for the test set.

A detailed examination of the actual minimized structures sheds more light on how the induced fit simulations help to increase the observed number of true positives. The minimized structure of mexiletine docked into the 2D6 (PDB code: 2F9Q) is illustrated in Figure 7. Both Figure 7(a) and (b) show the docked poses for the true positives. In Figure 7(a), the candidate has a better reactivity score than the one in Figure 7(b). However, by taking into account the induced fit effect, the candidate in Figure 7(b) has a lower energy as indicated by the three salt bridges between the protein and the ligands. Therefore, both of these candidates are picked as positives.

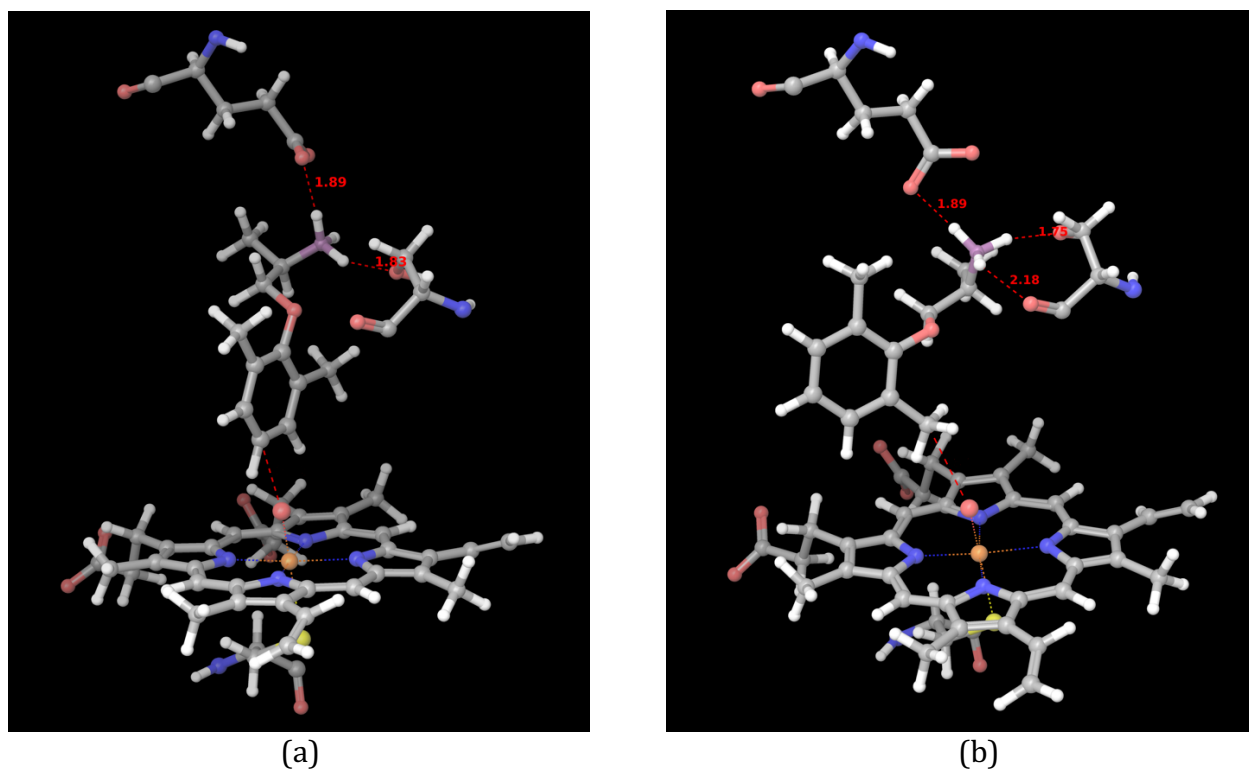

(a) (b)

Figure 7: The docked pose of mexiletine in P450 2D6. (a) pose for positive with better reactivity score. (b) pose for positive with worse reactivity score. The red dotted line with number shows the connection of the formed salt bridge between the ligand and protein. The red dashed line without number connects the oxo and target atom of the ligand. Color scheme is the same as that in Figure 5 with blue purple representing the positively charged nitrogen.

# 5. Conclusions

In the current study, we present a novel method to accurately and efficiently predict the sites of metabolism in reactions of drug-like molecules with the P450 isoforms 3A4 and 2D6. The multiple stage method that we have developed allows a flexible and accurate incorporation of both the intrinsic reactivity as well as the suitability of the ligand conformation for reaction, including induced fit effects. The contribution of the docking/induced fit component to the prediction accuracy is substantially larger in the case of 2D6, where the relatively small active site and presence of a salt bridge constraint allow robust structural prediction, than for 3A4, where the large, highly flexible active site presents severe challenges for current flexible docking methods; nevertheless, significant benefit to the docking component is seen even for the 3A4 case. For both isoforms, improvements as compared to purely ligand based methods are observed, with the largest improvement unsurprisingly being obtained for 2D6. The computation time is such that deployment at various stages of structure based drug design projects should be routine.

The results for 2D6 provide, in these retrospective studies, highly precise predictions. Further validation of the method requires one or more of the following; (a) prospective prediction of SOMs, confirmed by experiment; (b) confirmation of the structural models, which is possible via NMR experiments; (c) evaluation using larger and more diverse data sets. If the results hold up to these tests, the 2D6 model can be considered a useful practical solution to the prediction of 2D6 SOMs.

As discussed above, the 3A4 model will require significantly more work to get to this point. The key here is improving the structural prediction methodology for 3A4. Experimental information concerning 3A4 binding modes (again possibly obtained from NMR measurements) would be of great help in achieving this goal. However, better sampling algorithms and scoring functions may also be required. The task is a challenging one and it will require significant effort to make progress; the 2D6 results provide a target level of performance that one could hope to reach with a better protocol.

Finally, we intend to extend the present approach to a complete suite of P450 isoforms, including other significant enzymes such as 2C9. The remaining isoforms are more like 2D6 in that the active sites are not as larger as 3A4, and the binding motifs are better defined. Thus, we are optimistic that performance similar to that for 2D6 can be achieved for these additional isoforms. If this is the case, the overall improvement across all isoforms for SOM prediction will be significant, and can potentially make a valuable contribution to drug discovery projects in which modification of metabolic behavior is an issue.

# Supporting Information

| Type | Functional groups |
|---|---|
| **Primary** | C ( further classified by aromaticity, number of connected H and certain ring structure ) |
| | N ( further classified by aromaticity and number of connected H ) |
| | S (oxidation number < +4, further classified by aromaticity) |
| | P (bonded with S) |
| **Secondary** | -CH3 (contribution depends on the number of CH3) |
| | -F |
| | N-C=O (contribution depends on the connected primary groups) |
| | -O- (contribution depends on the number of O) |
| | Nonaromatic 6/5-member ring with O |
| | 6/5-member ring with one carbon double bonded to O |
| | Benzisoxazole-like structure |
| | N in aromatic ring |
| | C=C |
| | -N- (not in conjugated system); |
| | N-N |
| | C=C-C=O |
| | -S- |
| | N-C=S |
| | N-C≡N |

|  | N-C=N |
|---|---|
|  | 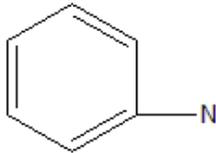 |
|  | 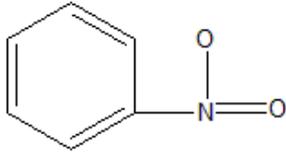 |
|  | 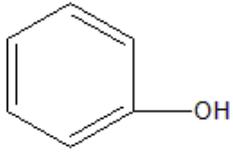 |
|  | 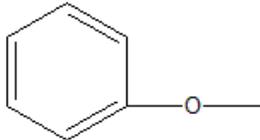 (contribution depends on the connected primary groups) |
|  | 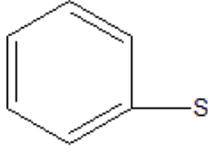 |
|  | 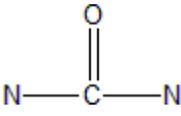 |
|  | 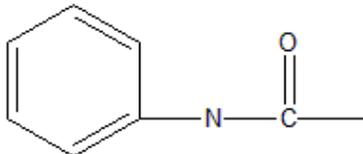 (contribution depends on the connected primary groups) |
| **Correctional** | C=O ( contribution depends on the connected primary and secondary groups) |

-O-

-OH

C=C

O-C=O ( contribution depends on the connected primary and secondary groups)

-Cl

-CH3

-N- (not in conjugated system)

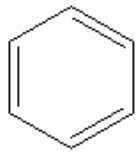

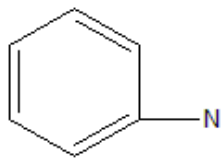

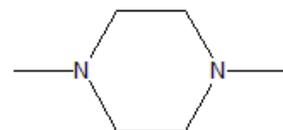
(contribution depends on whether the primary group is in the ring)

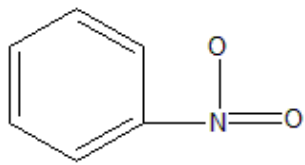

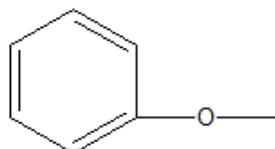
(contribution depends on the connected primary groups)

| | 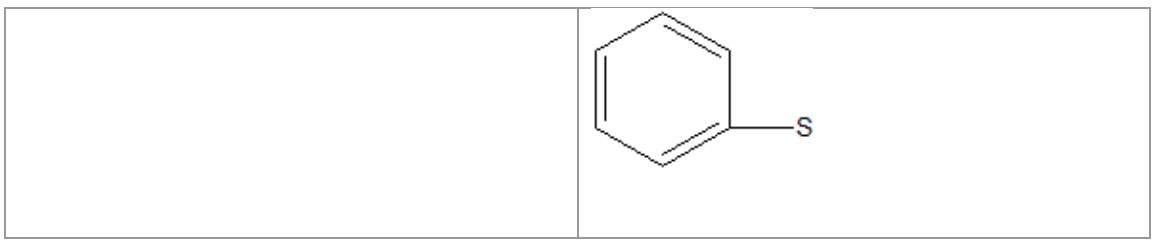 |
|---|---|

Table S1, List of functional groups used in the case of 3A4.